**Superconducting proximity effect in a transparent van der Waals superconductor-metal junction**

Jing Li[1†], Han-Bing Leng[2], Hailong Fu[1], Kenji Watanabe[3], Takashi Taniguchi[3], Xin Liu[2], Chao-Xing Liu[1*], Jun Zhu[1,4*]

[1]Department of Physics, The Pennsylvania State University, University Park, Pennsylvania 16802, USA.

[2]School of Physics and Wuhan National High Magnetic Field Center, Huazhong University of Science and Technology, Wuhan, Hubei 430074, China.

[3]National Institute for Material Science, 1-1 Namiki, Tsukuba 305-0044, Japan.

[4]Center for 2-Dimensional and Layered Materials, The Pennsylvania State University, University Park, Pennsylvania 16802, USA.

We report on Andreev reflections at clean $NbSe_2$-bilayer graphene junctions. The high transparency of the junction, which manifests as a large conductance enhancement of up to 1.8, enables us to see clear evidence of a proximity-induced superconducting gap in bilayer graphene and two Andreev reflections through a vertical $NbSe_2$-graphene and a lateral graphene-graphene junction respectively. Quantum transport simulations capture the complexity of the experimental data and illuminate the impact of various microscopic parameters on the transmission of the junction. Our work establishes the practice and understanding of an all-van-der-Waals, high-performance superconducting junction. The realization of a highly transparent proximized graphene – graphene junction opens up possibilities to engineer emergent quantum phenomena.

## I. Introduction

The proximity effect has been a central subject of superconductivity research for decades. Superconductivity correlation is introduced to the normal side of a superconductor (S)-normal metal (N) junction through a process known as the Andreev reflection (AR), where an incident electron is reflected back as a hole [1-3]. Current condensed matter research exploits the superconducting proximity effect to engineer exotic interfacial quantum phenomena such as topological superconductivity, for which a highly transparent S-N junction is critically important [4,5]. The ever expanding family of van der Waals metals, superconductors, topological insulators, and ferromagnets [6-12] makes a compelling case to explore the proximity effect in van der Waals S-N junctions, where advanced transfer techniques can produce sharp and clean interfaces. Graphene exhibits the proximity effect with a number of superconductors including a van der Waals superconductor $NbSe_2$ [13-27]. The specular Andreev reflection, which is unique to gapless materials, was observed [21,22,24]. Micrometer-scale transport of supercurrent has been reported in highly transparent Josephson junctions using conventional superconductors, such as MoRe [13-20]. However, prior $NbSe_2$-graphene junctions were considerably less transparent [21,22,25]. Improving on the quality of two-dimensional (2D)-2D S-N junctions and understanding the AR process in this unconventional geometry are essential steps to realize the potential of the van der Waals platform in illuminating fundamental quantum phenomena occurring at interfaces.

[†]Present address: National High Magnetic Field Laboratory, Los Alamos, NM 87544, USA

[*]Correspondence to: jzhu@phys.psu.edu (J. Zhu), cxl56@psu.edu (C.X. Liu)

In a 2D-2D S-N junction, the "interface" is an area where the S and N materials overlap and carriers tunnel in between. The normal component is often a semimetal or semiconductor with a gate-tunable carrier density, the magnitude and spatial distribution of which near the junction is expected to affect the AR process. The density-dependent carrier mean free path of the normal component plays a key role in establishing the superconductivity correlation [17,18,20]. In a 2D-2D S-N junction, the characteristic length scales of the carrier density profile, the mean free path and the device dimensions can all become comparable. A microscopic understanding of the AR process in such a system requires realistic modeling beyond the Blonder-Tinkham-Klapwijk (BTK) model established in traditional 3D metal S-N junctions [1,21,22,25].

In this article, we present electrical transport studies of ultra-transparent $NbSe_2$-bilayer graphene (BLG) S-N junctions, where the zero-bias differential conductance is enhanced by a factor of 1.8 due to ballistic AR. We report evidence of a proximity-induced superconducting gap in the BLG region directly bonded to the $NbSe_2$ and two AR processes occurring respectively at the vertical $NbSe_2$–BLG junction and the lateral proximitized BLG–normal BLG junction. Quantum transport simulations provide an excellent description of data and a microscopic understanding of the impact of the various parameters of the junction on the proximity effect.

## II. Experimental details

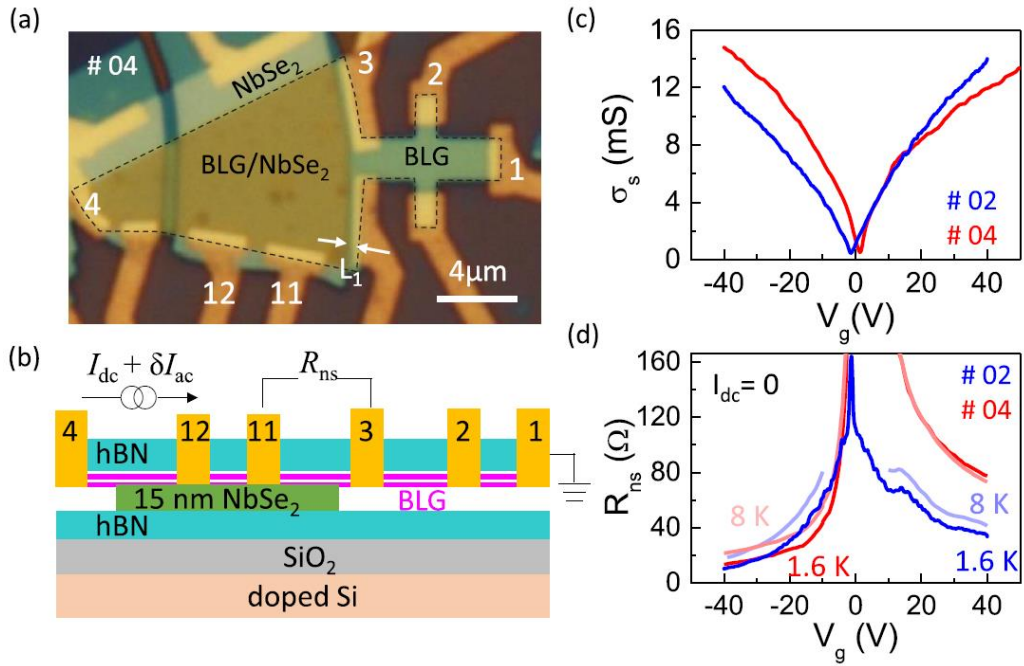

FIG.1. (a) An optical micrograph of device 04. $NbSe_2$ only, BLG/$NbSe_2$ and BLG only regions are labeled in the image. The dashed black line outlines the BLG sheet. $L_1 \approx 300$ nm marks the distance between the $NbSe_2$-BLG interface and the voltage probe. (b) A schematic side view of device 04. The differential junction resistance $R_{ns}$ ($I_{dc}$) i.e. $dV_{ac}/dI_{ac}$, is measured by passing a dc and a perturbative ac current from electrodes 4 to 1 and measuring the ac voltage between electrodes 11 and 3. Here the four-probe, current bias mode is chosen to avoid including the contact resistances, which far exceed the small $R_{ns}$ in our devices. (c) Sheet conductance of the BLG region vs. silicon gate voltage $V_g$ in device 02 (blue trace) and device 04 (red trace). $T$ = 1.6 K. Measurements are done in the BLG only region. (d) Zero-bias $R_{ns}$ ($V_g$) in devices 02 and 04 at $T$ = 1.6 and 8 K as labeled.

Our NbSe$_2$-BLG junctions are made by dry van der Waals transfer methods and encapsulated in hexagonal Boron Nitride (h-BN) sheets [28,29] (See Appendix A). The NbSe$_2$-BLG interface is free of polymer contaminant and the two sheets bond strongly during the transfer process. These are crucial elements to achieving a highly transparent S-N interface described below. In stacking order, the BLG is above/below the NbSe$_2$ respectively in device 04/02. Figures 1(a) and 1(b) show an optical image and sideview schematic of device 04. Figure 5 show similar depictions of device 02.

Transport measurements were carried out in a pumped He$^4$ cryostat at temperature $T = 1.6$ K unless otherwise noted. As shown in Fig. 1(b), we apply a variable dc current $I_{dc}$ together with a small ac current $\delta I_{ac}$ (200 nA, 17Hz) through the entire device from the NbSe$_2$ side (electrode 4) to the BLG side (electrode 1), and measure the differential resistance $R \equiv dV_{ac}/dI_{ac}$ as a function of $I_{dc}$ using a lock-in between different voltage probes. $R_{11,3}$ denotes the differential resistance measured between electrodes 11 and 3 across the junction, which we call $R_{ns}$. In the courses of the analysis, we found that plots of $R_{ns}$ vs $I_{dc}$ are most instructive. The conventional $dI/dV$ vs $V_{dc}$ plot $G_{ns}$ ($V_{dc}$) is computed from the measured $R_{ns}$ ($I_{dc}$) data through integration then differentiation. Results on device 04 are shown in Fig. 7 to show their relations and facilitate comparison to other systems. NbSe$_2$ sheets used in this study (~15-nm-thick in device 04 and ~10-nm-thick in device 02) exhibit the superconducting behavior similar to that of bulk NbSe$_2$ [30] with a critical temperature $T_c$ of 7.0 K (see Fig. 5). Figure 1(c) plots the back-gate $V_g$-dependent sheet conductance $\sigma_s$ ($V_g$) on the BLG side of both device 04 and 02, respectively. In device 04, the carrier Hall mobility at $V_g = \pm 40$ V is $\mu_{Hall}$ = 28 000 and 32 000 cm$^2$/Vs, which corresponds to mean free path $l_{mfp} = \hbar\mu\sqrt{\pi n}/e$ = 780 and 930 nm respectively. Device 02 has comparable quality. In both devices $l_{mfp}$ is a few times larger than the distance between the voltage probe and the NbSe$_2$-BLG interface $L_1$ marked in Fig. 1(b). This ensures that an Andreev reflected hole can travel ballistically to the voltage probe before being scattered.

Figure 1(d) compares junction differential resistance $R_{ns}$ ($V_g$) in the superconducting ($T = 1.6$ K, darker traces) and normal state ($T = 8$ K, lighter traces) of the NbSe$_2$ in both devices. Here $I_{dc} = 0$. Device 04 exhibits pronounced *e-h* asymmetry. This is not surprising as charge transfer from NbSe$_2$ [18,20] pins the left side of the BLG in the hole regime and a *p-n/ p-p* junction forms when $V_g$ is positive/negative. The asymmetry is much smaller in device 02 as $V_g$ dopes both sides; the difference arises from the opposite stacking order of the NbSe$_2$ and BLG sheets. In the normal state, $R_{ns}$ drops rapidly with increasing doping and reaches 21 Ω / 18 Ω, respectively, in device 04/ 02 at $V_g = -40$ V. This low $R_{ns}$ is on par with the best elemental metal superconductor-graphene junction resistance reported in the literature [20] and much smaller than the hundreds to thousands of Ω obtained in previous NbSe$_2$-graphene junctions [21,22,25]. The high interface transparency is key to our observation of the proximity-induced gap in the BLG and the second AR at the proximitized BLG-normal BLG lateral junction.

### III. Results and discussion

Figures 2(a) and (b) plot $R_{ns}$ ($I_{dc}$) in device 04 at fixed $V_g = +40$ V (electron) and -40 V (hole) respectively for a set of temperatures ranging 1.6 - 8 K. $R_{ns}$ ($T$) in the central $I_{dc}$ region of approximately -60 μA < $I_{dc}$ < 60 μA starts to deviate significantly from its normal state values below the $T_c$ of NbSe$_2$. Outside this region and up to a few hundred μA, $R_{ns}$ (1.6 K) ≈ $R_{ns}$ (8 K) apart from a few small resistance spikes that suggest local heating hot spots in the NbSe$_2$ sheet [14] (See Appendix B for details). In the bias range of 60 μA > |$I_{dc}$| > 20 μA, both electron and hole data show increasing reduction of $R_{ns}$ with decreasing $T$.

The reduction of $R_{ns}$ is similar in magnitude despite their very different normal state resistances. At lower current biases, sharp resistance spikes develop in the electron regime [onset marked by open circles in Fig. 2(a)], while a curvature change appears at similar $I_{dc}$'s on the hole side but $R_{ns}$ continues to decrease [Fig. 2(b)].

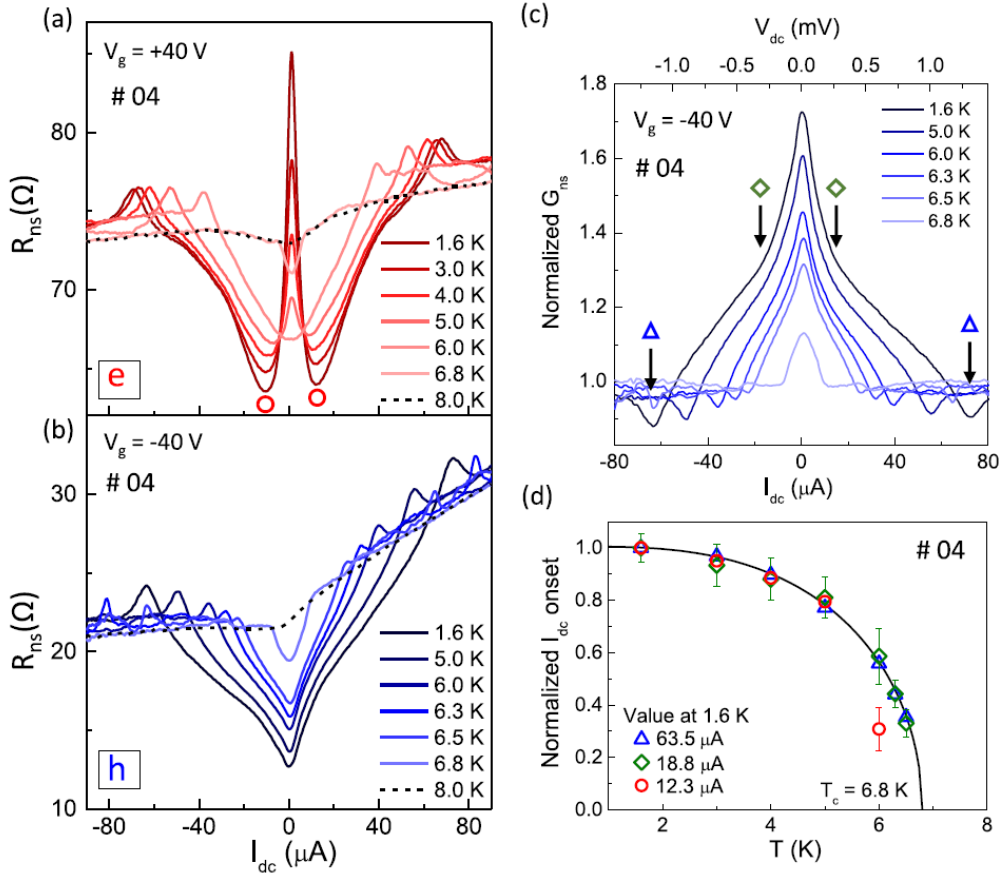

FIG. 2. $R_{ns}$ ($I_{dc}$) for $V_g$ = +40 V (a) and -40 V (b) at selected temperatures as labeled in the plots. Red circles in (a) mark the onset of the sharp resistance spike for the 1.6 K trace. (c) Normalized $G_{ns}$ ($I_{dc}$) at $V_g$ = -40 V using data in (b). The top axis plots the voltage drop across the junction, obtained through integration. Blue triangles and green diamonds mark the onset of the dome and zero-bias peak respectively for the 1.6 K trace. (d) $T$-dependent $I_{dc}$'s of the circles, triangles and diamonds marked in (a) and (c), averaged between positive and negative onsets and normalized to their 1.6 K value. Black curve is a fit to the BCS gap model: $I_{dc} \propto \tanh\left(1.74\sqrt{\frac{T_c}{T} - 1}\right)$, where $T_c$ = 6.8 K. From device 04.

Figure 2(c) shows the normalized conductance $G_{ns}$ ($T$) = $R_{ns}$ (8 K)/ $R_{ns}$ ($T$) vs $I_{dc}$ for the hole data in Fig. 2(b). The normalized $G_{ns}$ ($T$) vs $V_{dc}$ conventional plots are nearly identical in shape (Fig. 7). $G_{ns}$ exhibits a pronounced "dome plus peak" structure, with the onset of the dome (marked by triangles) occurring at $I_{dc} \sim \pm 63.5$ μA ($V_{dc} \sim 1.2$ meV), which agrees very well with the superconducting gap $\Delta_0 = 1.2$ meV in NbSe$_2$ [31]. The inner $G_{ns}$ peak onsets at ~ 20 μA marked by the diamonds and reaches 1.75 times the normal state value at $I_{dc} = 0$. In a similar bias range marked by the circles in Fig. 2(a), the electron resistance/conductance data exhibits pronounced peaks/dips [Fig. 7(e)]. We track the temperature dependence of the triangles, diamonds and circles and plot all three in Fig. 2(d). A fit to the $T$ dependence of a Bardeen-Cooper-Schrieffer (BCS) superconducting gap $\Delta(T) \propto \tanh\left(1.74\sqrt{\frac{T_c}{T} - 1}\right)$ [32] with $T_c$ = 6.8K [solid black line in Fig. 2(d)] describes all three symbols very well. This analysis suggests that all three thresholds are related to the superconducting gap $\Delta_0$ in NbSe$_2$ and are thus proximity-induced in origin, as opposed to coming from an intrinsic second gap in NbSe$_2$ [25,33-35]. The circles and diamonds correspond to energies of 0.2$\Delta_0$ and 0.3$\Delta_0$, respectively.

A 100% efficient AR converts all incident electrons to holes and thus increases the junction conductance twofold [3]. In real materials, significant enhancement of conductance is only seen in highly transparent [36,37] or highly disordered [38] S-N junctions. In past graphene S-N junctions, the excess conductance is typically only a few percent [21,22,25]. Here, the zero-bias conductance enhancement factor of 1.75 is quite remarkable and together with a very small $R_{ns}$, confirms the highly transparent interface we achieved. It is only in these clean devices that a proximity-induced gap $\Delta_1 \sim$ 0.2-0.3 $\Delta_0$ is revealed. As illustrated in Fig. 3(a), we hypothesize that $\Delta_1 = \Delta_{BLG}$ represents a proximity-induced gap of the BLG region in direct contact with the NbSe$_2$. In this scenario, for electrons and holes in the energy range of $\Delta_0 > |E| > \Delta_{BLG}$, this BLG region is normal and AR occurs at the vertical NbSe$_2$-BLG interface. This gives rise to resistance reductions that are approximately independent of $V_g$ in device 04 because the BLG/NbSe$_2$ region is not affected by the back-gate. This is indeed what our data in Figs. 2(a) and 2(b) showed. A lateral superconducting -normal BLG junction emerges for carriers of energy $0 < |E| < \Delta_{BLG}$. This lateral junction is accompanied by a carrier density change, as illustrated in Fig. 3(b) for $V_g = \pm 40$ V (see Fig. 9 for COMSOL simulations). The presence of a $p$-$n$ junction at positive $V_g$ increases the normal backscattering amplitude for electrons and suppresses AR. A $p$-$p'$ junction, which is much more transparent, forms at negative $V_g$ and promotes AR. This $V_g$-dependent carrier density profile can thus account for the $e$-$h$ contrasting peak/dip seen in the low bias range of $R_{ns}$ in Figs. 2(a) and 2(b).

A comprehensive quantum transport simulation enables us to confirm the above physical picture and obtain microscopic insights of the two ARs in this 2D-2D S-N junction. Our simplified two-Andreev model illustrated in Fig. 3(c) consists of three regions NbSe$_2$ (I, $\Delta_0$), proximitized BLG (II, $\Delta_{BLG}$), and normal BLG (III), which are connected by two scattering interfaces A and B. Our simulations show that the presence of the proximity-induced $\Delta_{BLG}$ and the formation of a $p$-$p$ or $p$-$n$ junction at the NbSe$_2$-BLG interface are essential for the appearance of an inner conductance peak/dip for hole/electron carriers in experiments. These features remain robust over a wide range of parameters and scenarios tested in the calculations (Appendix F) with no artificial barriers or reflection coefficient added. Briefly, interface A represents the *vertical* tunnel junction between NbSe$_2$ and BLG with a linear gap variation over a small representative width $L_A = x_1 - x_0 = 5a$, where $a$ is the lattice constant of graphene. The barrier strength at interface A is set to zero in the calculation, reflecting the highly transparent NbSe$_2$-BLG interface. In a real device tunneling can occur at a variable distance from the physical NbSe$_2$-BLG boundary. This is modeled by varying the width of region II, $L_{II} = x_2 - x_1$ in our calculations (Fig. 12). Interface B represents the *lateral* superconducting-normal BLG junction with a representative width of $L_B = x_3 - x_2 = 20a$. $\Delta_{BLG}$ ($x$) decays linearly within $L_B$. To model device 04, we set the chemical potential in BLG $\mu$ ($x$) is a constant in regions I and II and takes on different values in region III depending on the doping level. For simplicity, we have chosen $\mu$ ($x$) and $\Delta_{BLG}$ ($x$) to have the same functional form in Fig. 3(c). Our simulations (Fig. 12) show that the underlying physics is not sensitive to the specific choices shown here as long as $\Delta_{BLG}(x)$ varies smoothly and $\mu(x)$ varies over the same region or extends further into the normal BLG region, which are likely the case in real devices.

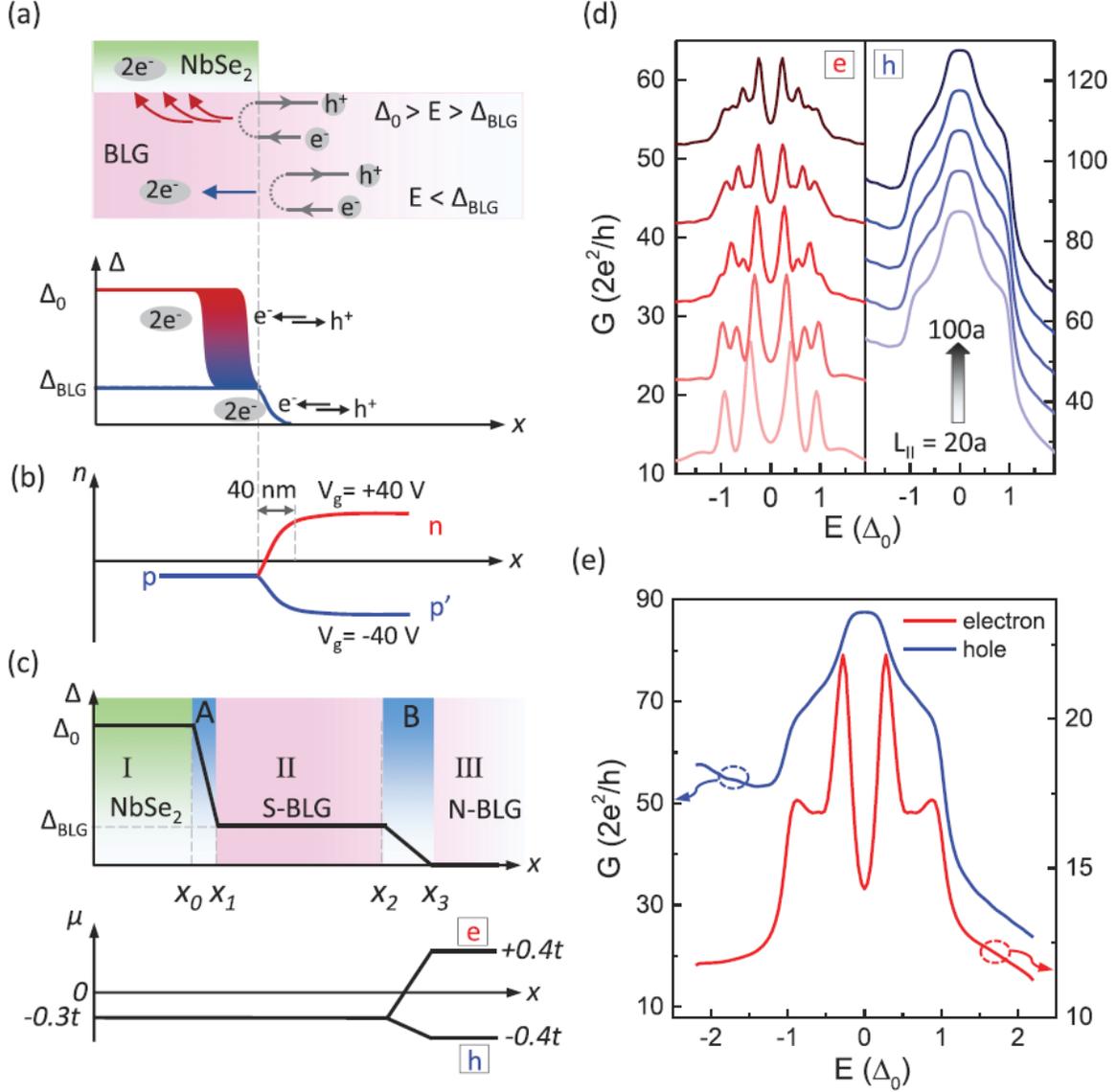

FIG. 3. (a) Illustrations of the two gaps and two AR processes in our NbSe$_2$-BLG junction. $\Delta_0$ is the superconductor gap of NbSe$_2$. $\Delta_{BLG}$ is the proximity-induced gap in the BLG region in direct contact with the NbSe$_2$ sheet. In device 04, the BLG is above the NbSe$_2$. (b) COMSOL simulated carrier density profile $n(x)$. (c) The three-region model and the profile of the chemical potential $\mu$ in BLG used in our simulations. $t$ is the in-plane nearest neighbor hopping energy in graphene. (d) Simulated two-terminal conductance $G(E)$ at selected $L_\parallel$ values from $20a$ to $100a$ in $20a$ steps for both electron ($\mu = 0.4t$, left panel) and hole ($\mu = -0.4t$, right panel) doping. Curves are vertically shifted by 10 units for clarity. Parameters used are $\Delta_0 = 0.05t$, $k_BT = 0.04\Delta_0$. $\Delta_{BLG} = 0.2\Delta_0$. (e) $G(E)$ obtained by averaging the curves in (e).

We numerically compute the two-terminal conductance of the junction $G$ using a generalized Landauer-Büttiker formalism implemented in the Kwant program [39] (See Appendix F). Many microscopic scenarios are explored and the details are given in Secs. 2-5 of Appendix F. Here Fig. 3(d) plots an example of $G(E)$ for electron (left panel) and hole (right panel) doping regime, respectively, for different $L_{II}$'s ranging from $20a$ to $100a$. It is clear that the main features of the data, i.e. enhanced $G(E)$ inside $\pm\Delta_0$ and the contrasting behavior of electrons and holes at $E < \Delta_{BLG}$ are reproduced by the calculation. Simulations of the electron regime show conductance oscillations correlated with the length of $L_{II}$. They are Fabry-Perot interference effects resulting from Andreev reflections at interface A and normal reflections at interface B (see Fig. 12). No such oscillation is seen in the hole regime, where interface B is much more transparent. These oscillations also did not appear in measurements since $L_{II}$ varies in *real* devices. To mimic experiment, we have averaged $G(E)$ of the curves shown in Fig. 3(d) and plotted the results in Fig. 3(e). Figure 3(e) reproduces very well the important features of our data. Notably, no disorder scattering is included in our simulations, thus the good agreement between theory and experiment verifies the ballistic transport nature of our NbSe$_2$-BLG junction.

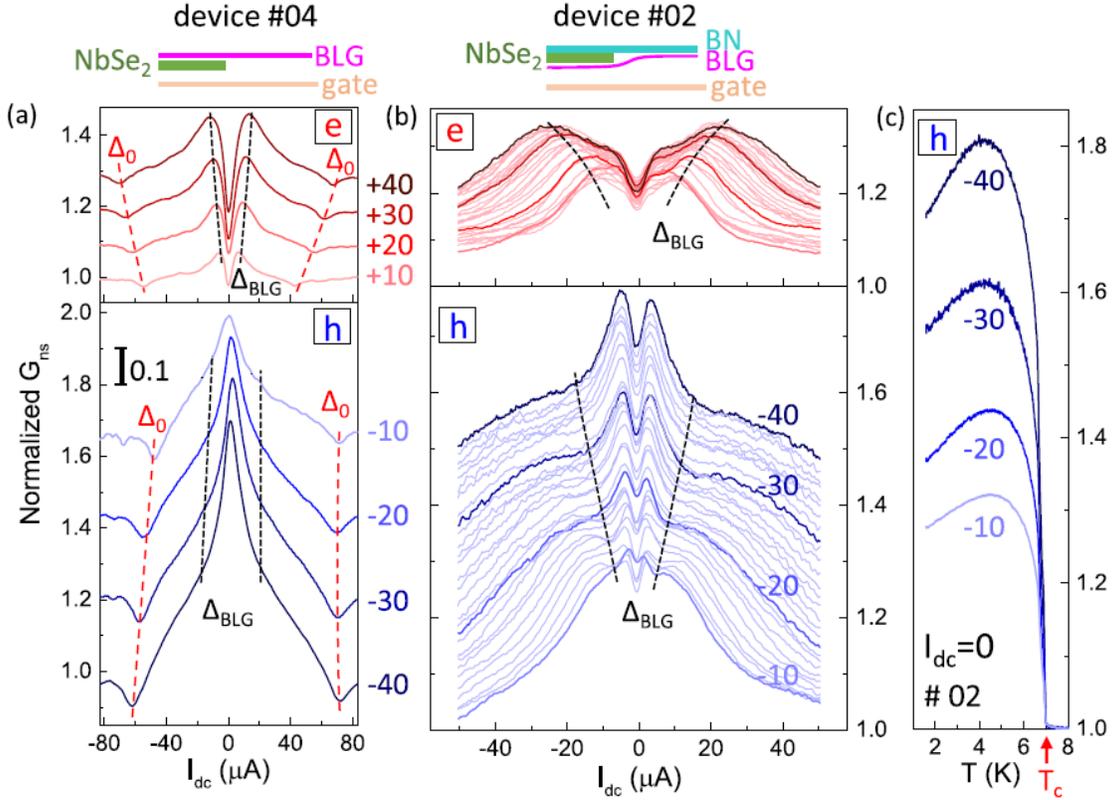

FIG. 4. Normalized $G_{ns}(I_{dc})$ in device 04 (a) and 02 (b) at selected $V_g$'s as labeled in the plots. Curves in (a) are vertically shifted by 0.1 (upper panel) and 0.2 (lower panel) for clarity except for the bottom curve. Curves in (b) are plotted as is. From top to bottom: $V_g$ changes from +40 V to +10 V (upper panel), and from -40 V to -10 V (lower panel) in step of 1 V. Upper and lower panels in each figure share the same y-scale. Red and black dashed lines are guide to the eye showing the positions of $\Delta_0$ and $\Delta_{BLG}$ respectively. The schematics illustrate the different gating situations in devices 04 and 02. $T = 1.6$ K. (c) Temperature dependence of the zero-bias $G_{ns}$ in device 02. From top to bottom: $V_g = -40, -30, -20$ and $-10$ V.

Measurements in a second device 02, and the $V_g$ and $T$ dependence of the normalized $G_{ns}$ further support the two-AR junction model we established. Figures 4(a) and 4(b) compare the normalized $G_{ns}(I_{dc})$ in both devices at selected $V_g$'s in the electron (top panels) and hole (bottom panels) doping regimes, where we have marked the trend lines of $\Delta_0$ and $\Delta_{BLG}$. Similar to device 04, $G_{ns}$ in device 02 is enhanced by the onset of superconductivity–a full set of $T$-dependent data extending the current range to $\Delta_0$ are given in Fig. 8–and exhibits carrier-specific behaviors at $|E| < \Delta_{BLG}$ as Fig. 4(b) shows. The different gating situations have also led to some important differences in the two devices. Device 02 exhibits a weaker electron-hole asymmetry in both the normal and superconducting state of $R_{ns}$ because the back-gate acts on both sides of the BLG. Also, unlike device 04, the $G_{ns}$ ($I_{dc}$) of device 02 develops a small zero-bias dip at $T < 4$ K even for hole carriers [Fig. 4(b) bottom panel], which suggests the presence of a small barrier at the lateral BLG-BLG junction. We suspect that a stacking-induced curvature of the BLG sheet as shown in the diagram above the graph might play a role. Finally, we show in Fig. 4(c) the temperature dependence of the normalized junction conductance $G_{ns}$ ($T$, $I_{dc} = 0$) in device 02. $G_{ns}$ ($T$, $I_{dc} = 0$) rises sharply at $T < T_c$ and reaches values as high as 1.8 before decreasing slightly at low temperatures. Its behavior agrees well with the $T$-dependence of Andreev reflections reported in the literature [37].

## IV. Conclusions

In conclusion, we have fabricated and studied very transparent $NbSe_2$-BLG S-N junctions, where ballistic Andreev reflections give rise to a conductance enhancement of up to 1.8 in the superconducting state of our devices. Experiment and theory show that the transmission across the junction undergoes two different Andreev reflections at low and high energies. These insights, only revealed in our high-quality devices, are expected to be also relevant to other types of van der Waals superconducting devices. The attainment of an ultra-transparent lateral S-N junction within the same BLG sheet offers a pathway to construct high-quality superconducting devices.

*Note added*: While our manuscript was under review, we became aware of a related study by Moriya et al on $NbSe_2$-graphene junctions, which also concluded on the occurrence of two Andreev reflection processes in their devices [40].

## Acknowledgement

J. L., H.-B. L. and H. F. contributed equally to this work. Work at Penn State is supported by the NSF through NSF-DMR-1708972 and Kaufman New Initiative research grant KA2018-98553 of the Pittsburgh Foundation. H. F. acknowledges the support of the Penn State Postdoctoral Eberly Fellowship. C.X. L. acknowledges the support from the Office of Naval Research (Grant No. N00014-18-1-2793) and DOE grant (DE-SC0019064). X. L. and H.-B. L. acknowledge the support from the National Key R&D Program of China (Grant No. 2016YFA0401003) and NSFC (Grant No.11674114). K.W. and T.T. acknowledge support from the Elemental Strategy Initiative conducted by the MEXT, Japan and the CREST (JPMJCR15F3), JST.

**Appendix A: Device fabrication procedure and characteristics**

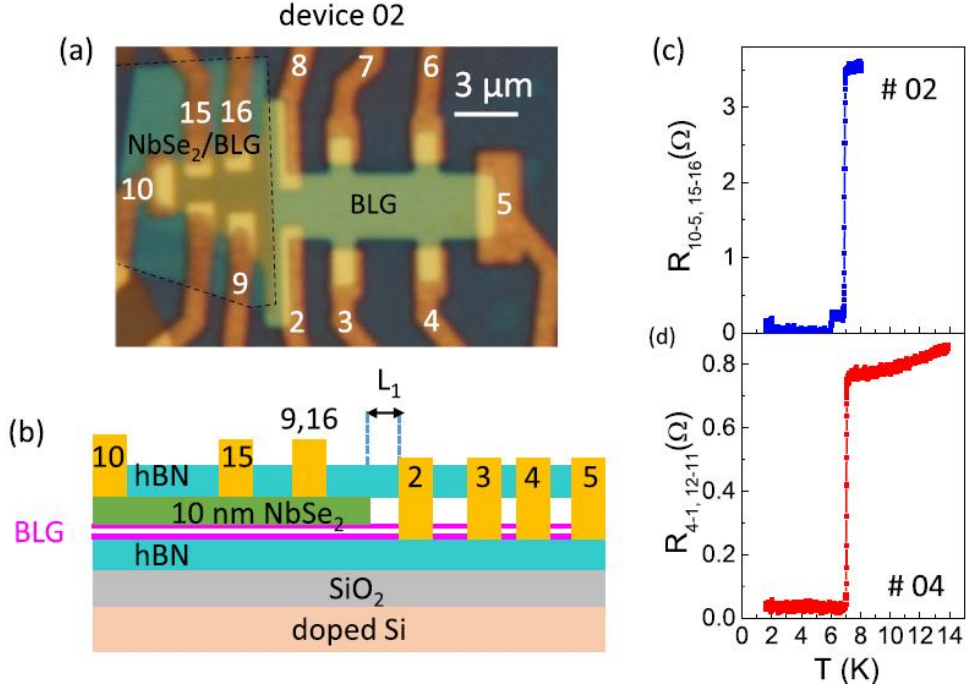

FIG. 5. (a) An optical micrograph of device 02. NbSe$_2$/BLG and BLG only regions are labeled in the image. The dashed black line outlines the NbSe$_2$ flake. (b) A schematic side view of device 02. $L_1 \approx$ 300 nm marks the distance between the NbSe$_2$-BLG interface and the voltage probe, which is ~ 450 nm in device 02. (c) and (d) plot temperature-dependent resistance measured on the NbSe$_2$/BLG side of both devices. $R_{a-b,c-d}$ denotes four-terminal measurement using current flow from electrodes a to b and measuring voltage between electrodes c and d. $T_C$ = 7.0 K in both measurements.

Our hBN encapsulated NbSe$_2$-BLG junctions are made by dry van der Waals transfer using a polypropylene carbonate (PPC) stamp. The majority of the process follows the established practice of stacking hBN and graphene [28,41]. Here, NbSe$_2$ flakes are exfoliated onto a SiO$_2$ wafer and picked up right away by an hBN flake (device 02) or an hBN/BLG stack (device 04). We put a portion of the NbSe$_2$ flake in contact with the PPC film to help lift it from the SiO$_2$ surface. After making contact, the PPC stamp is quickly lifted at a speed of 14 μm/s and around a temperature of 46 °C in the ambient. We etch the top hBN layer to expose the side or the top surface of the BLG or NbSe$_2$ flake to make Cr/Au contacts [28,41]. Figure 1(a) and Fig. 5(a) show the optical images of finished devices 04 and 02 respectively.

**Appendix B: Additional resistance spikes due to local heating effect**
Please see Fig. 6.

**Appendix C: Comparison of data in formats of $G_{ns}$ or $R_{ns}$ vs $I_{dc}$ or $V_{dc}$**
Please see Fig. 7.

**Appendix D: Additional data on device 02**
Please see Fig. 8.

**Appendix E: COMSOL simulation of carrier density profile**

Please see Fig. 9.

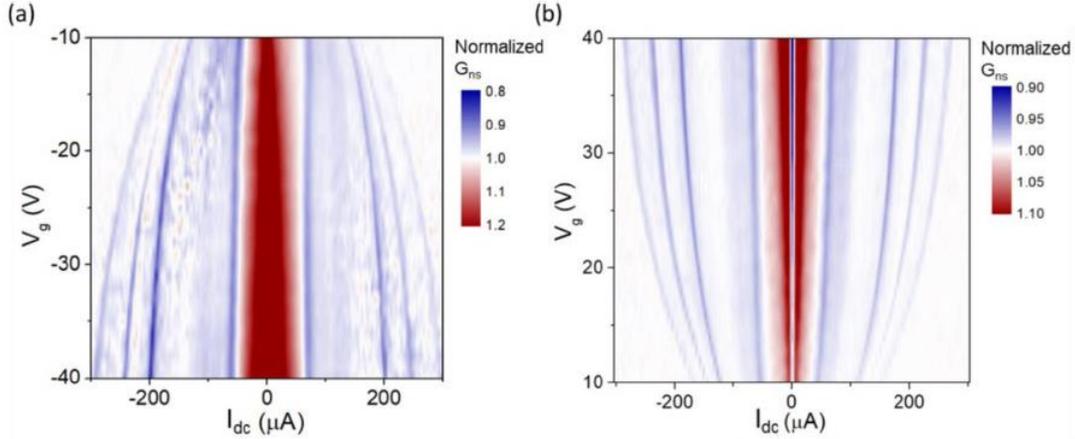

FIG. 6. A false color mapping of the normalized conductance $G_{ns} = R_{ns}$ (8 K)/ $R_{ns}$ (1.6 K) vs $I_{dc}$ and the back-gate voltage $V_g$ in the hole (a) and electron (b) doped regime in a large span of $I_{dc}$. From device 04. Here $R_{ns}$ is measured between electrodes 11 and 3 in Fig.1. Additional conductance dips (blue lines) are seen outside -60 μA < $I_{dc}$ < 60 μA. In measurements taken entirely on the BLG side, e.g. between electrodes 3 and 2 in Fig. 1, these conductance dips disappear while AR features inside the range -60 μA < $I_{dc}$ < 60 μA continue to manifest weakly thanks to the long ballistic length of our devices. These observations suggest that the conductance dips at large current are due to the loss of superconductivity in the NbSe$_2$ sheet at local hot spots generated by current heating.

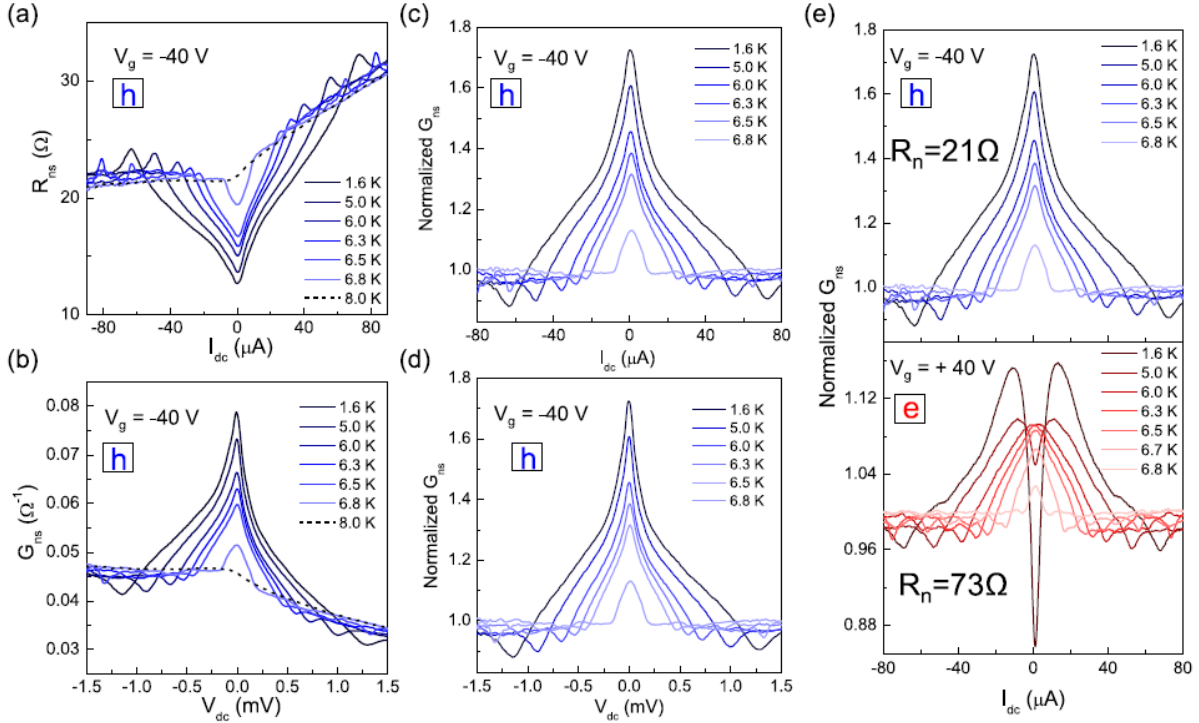

FIG. 7. (a) plots the differential resistance $R_{ns}$ ($I_{dc}$) in device 04 at selected temperatures as labeled in the plot. This is what we measured between electrodes 11 and 3. (b) plots the differential conductance $G_{ns}$ vs $V_{dc}$. $V_{dc}$ is computed by integrating measurements in (a). On the positive bias side, contributions to $V_{dc}$ from the nonlinear $R_{ns}$ ($I_{dc}$) background at 8 K is subtracted. The subtraction is about 0.2 mV at $I_{dc}$ = 50 μA, for example. (c) and (d) are normalized $G_{ns}$ vs $I_{dc}$ or $V_{dc}$ respectively. Comparing these plots, we see that all features of data are represented well on each plot and a threshold dc bias current of 60 μA × a normal state resistance of 20 Ω =

1.2 meV yields the expected gap Δ of NbSe$_2$. This indicates that all the dc voltage measured between 11 and 3 drops across the S-N junction. (e) plots the normalized $G_{ns}$ vs $I_{dc}$ at $V_g$ = -40 V (top panel) and + 40 V bottom panel). The conductance enhancement onsets at the same dc bias current in both. This suggests that the NbSe$_2$/BLG junction resistance is similar in both p- and n- doped regimes. The additional ~50 Ω resistance in the n- doped regime comes from the p-n junction. In this case, converting $I_{dc}$ to $V_{dc}$ would not capture the superconducting gap of NbSe$_2$ correctly. We have thus opted to plot data against $I_{dc}$ directly and further, use $R_{ns}$ instead of $G_{ns}$ to establish connections between the electron and hole regimes in Fig. 2.

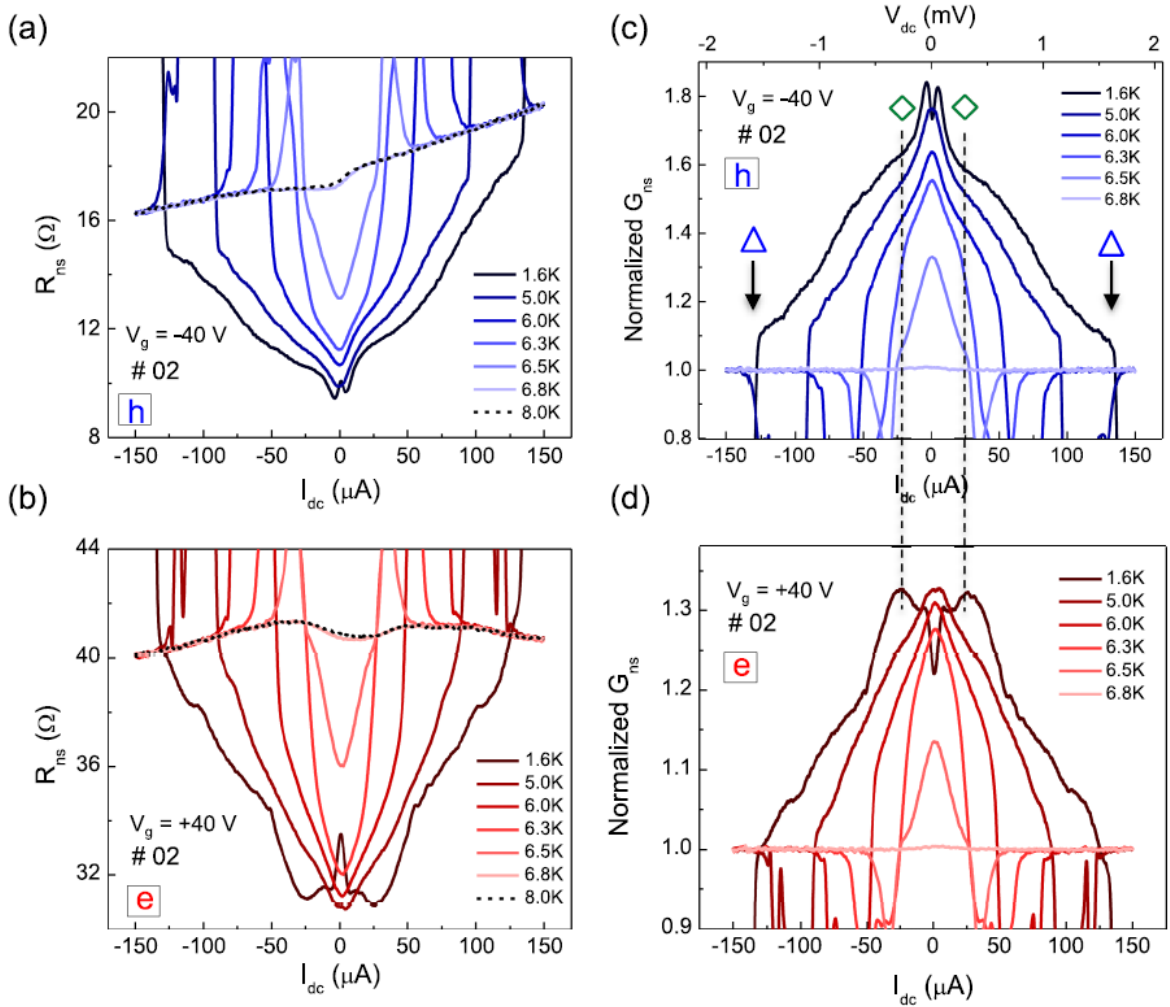

FIG. 8. Expanded $R_{ns}$ ($I_{dc}$) in device 02 as a function of temperature for $V_g$ = -40 V (a) and +40 V (b) plotted in a style similar to that of device 04 in Fig. 2. (c) and (d) plot normalized $G_{ns}$ ($I_{dc}$) obtained from data in (a) and (b) respectively. The top axis of (c) marks the computed voltage drop across the junction. The onset of enhanced conductance approximately corresponds to the superconducting gap of NbSe$_2$. Sudden resistance spikes near the threshold current are caused by the loss of superconductivity at the local hot spots in NbSe$_2$ due to larger dc current and thinner NbSe$_2$ flake used in this device. The dome plus inner peak/dip structure is similar to device 04, with a smaller difference between the two types of carriers observed. A small zero-bias conductance suppression develops at the lowest temperatures for both carriers, indicating a small tunnel barrier at the junction.

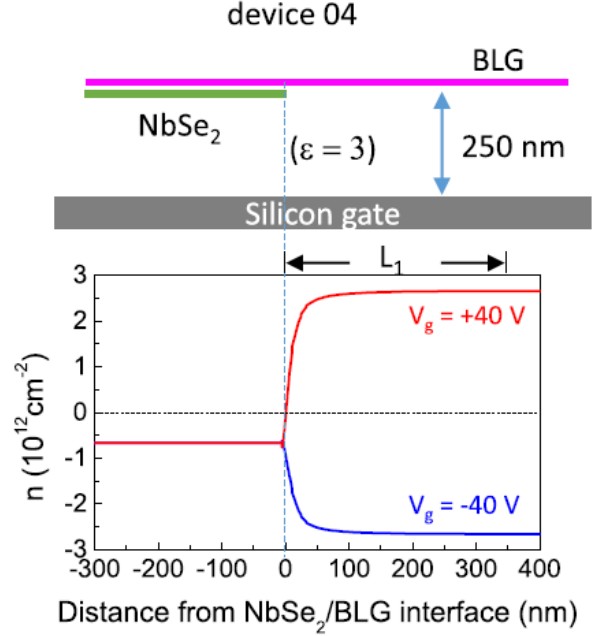

FIG. 9. (Upper panel) COMSOL setup used to simulate the carrier density profile in device 04. The NbSe$_2$ flake, the BLG flake and the silicon gate are each represented by 5-nm-thick metal slabs. The NbSe$_2$-BLG separation is 5 nm. To simplify the setup, we have set the BLG to silicon distance to 250 nm to represent the combined gating effect of a bottom hBN flake (~ 23 nm, $\varepsilon$ = 3) and a 295-nm-thick SiO$_2$ film ($\varepsilon$ = 3.9). The whole setup is immersed in a dielectric environment of $\varepsilon$ = 3 to represent the hBN encapsulation. The chemical potential on the NbSe$_2$ is fixed to a negative value to represent an experimentally informed charge transfer amount (hole type) to the BLG above. (Lower panel) Simulated carrier density profile $n(x)$ with silicon gate voltage $V_g$ = +40 V (red trace) and -40 V (blue trace). A *p-n* / *p-p* junction forms in the red/blue trace. The voltage probe is $L_1$ = 350 nm away from the NbSe$_2$-BLG interface.

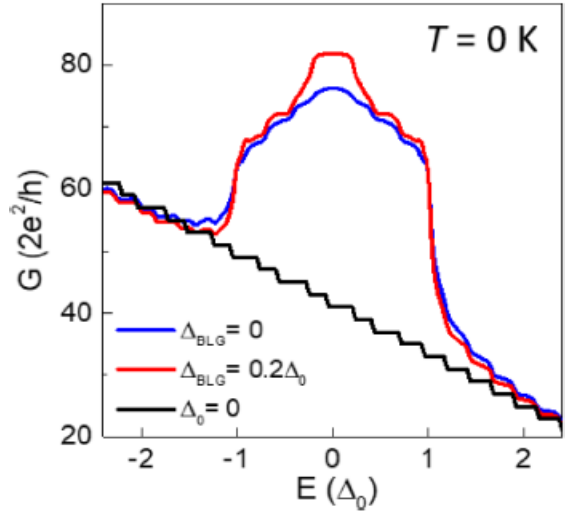

FIG. 10. (a) $G(E)$ computed for $\Delta_{BLG}$ = 0 (blue trace), $0.2\Delta_0$ (red trace) and the normal state (black trace) on the hole side. The dome + peak structure emerges when a nonzero $\Delta_{BLG}$ is used. $\Delta_0$ = 0.05$t$, $\mu_{BLG}$ = -0.3$t$, $L_{\parallel}$ = 50$a$. $L_B$ = 20$a$. $T$ = 0. The sloping background is caused by the changing number of modes in the bias window, as a consequence of the finite size effect.

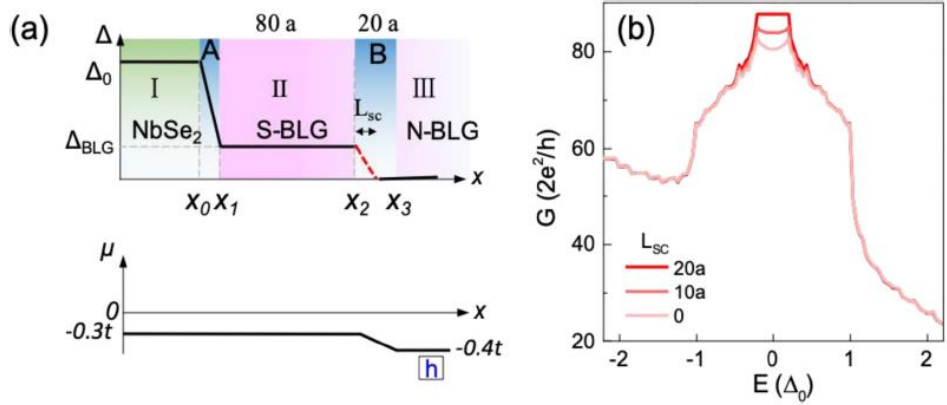

FIG. 11. (a) The spatial profile of Δ (x) and µ (x) used in simulations show in (b). (b) plots G (E) computed for $L_{sc}$ = 0, 10a and 20a respectively. $L_{II}$ = 80a, $L_B$=20a, T = 0, see Fig. 3 for other parameters.

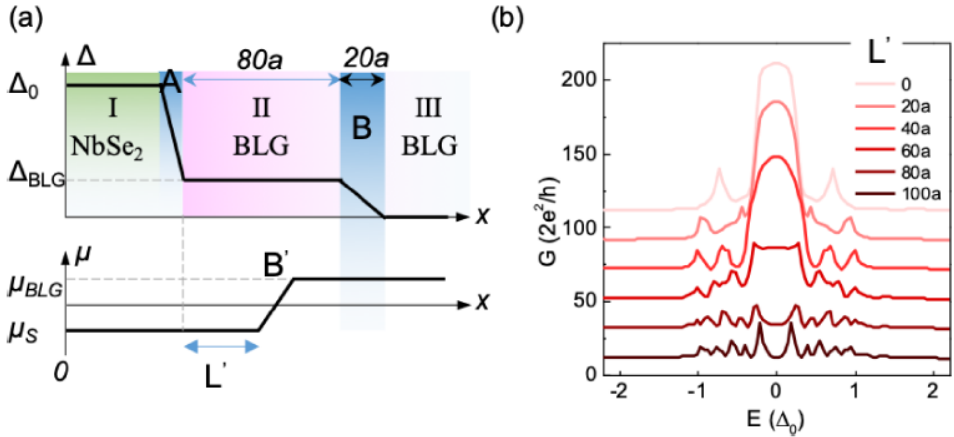

FIG. 12. (a) The spatial profile of Δ (x) and µ (x) used in simulations shown in (b). (b) plots the calculated G (E) in the electron doping regime with selected values of L' ranging from 0 to 100a. T = 0. See Fig. 3 for other parameters of the simulation. Curves are vertically shifted by 20 conductance units for clarity.

## Appendix F: Quantum transport model

### 1. Model Hamiltonian and numerical methods

We model BLG with a tight-binding Hamiltonian,

$$H_0 = -t \sum_{<i,j>,m} \left(a^\dagger_{m,i} b_{m,j} + H.c.\right) - t_\perp \sum_i \left(a^\dagger_{1,i} b_{2,i} + H.c.\right) \quad (S1)$$

where a and b are the electron annihilation operators, m = 1, 2 and i, j = A, B denotes the layer and sublattice index, respectively. t = 2.7 eV and $t_\perp$ = 0.4 eV are, respectively, the nearest neighbor intra- and inter- layer hopping energy. The superconductor (SC) proximity effect is introduced using the Bogoliubov de Gennes Hamiltonian,

$$H_{\text{BdG}} = \begin{pmatrix} H_0 - \mu(x) & \Delta(x) \\ \Delta^*(x) & -H_0^* + \mu(x) \end{pmatrix} \quad (S2)$$

where $\Delta$ is the SC gap, and $\mu$ is the relative chemical potential with respect to the charge neutrality point of the BLG. Both $\Delta$ and $\mu$ are spatial dependent functions along the direction of current flow $x$. In our convention, $\mu$ is $> 0$ for electron doping and $< 0$ for hole doping.

We divide the S-N junction into three regions, as illustrated in Fig. 3(c). Region I-III represents the NbSe$_2$, the superconducting BLG and the normal BLG areas respectively, which are connected by two interfaces A and B. For simplicity, we model NbSe$_2$ with the same Hamiltonian as the BLG (Eq. S2). Using experimental input, we take $\Delta = \Delta_0$ (SC gap of bulk NbSe$_2$) and $\Delta_{\text{BLG}} = 0.2\Delta_0$, 0 for regions I, II and III respectively, and model the decay of the SC gaps at the two interfaces with linear functions. We set $L_A = 5a$ and $L_B = 20a$, respectively to reflect the abrupt change of the SC gap from $\Delta_0$ to $\Delta_{\text{BLG}}$ at interface A and a more slowly decay of $\Delta_{\text{BLG}}$ at interface B. Features of the calculations are insensitive to the specific values of $L_A$ and $L_B$. In accord with the gating geometry of device 04, we set $\mu = \mu_s = -0.3t$ for regions I and II. In region III, $\mu = \mu_{\text{BLG}}$ depends on the silicon gate voltage. $\mu_{\text{BLG}} = +/- 0.4t$ is used to simulate electron/hole doping. The variation of $\mu$ at interface B is also approximated by a linear function.

In our numerical simulations, the graphene lattice constant $a = 1$ defines the unit of length. The sample width $W$ perpendicular to the current flow is set to $100a$. The finite sample size introduces a quantization energy scale on the order of $0.01t$, which manifests as steps in the conductance plots (e.g. Fig. 10). The majority of our simulations use an exaggerated $\Delta_0 = 0.05t$ to avoid the finite-size effect discussed in Fig. 13 while preserving the qualitative features of the model. We use the Kwant program [39] to calculate the Andreev reflection ($R^A$) coefficients at interfaces A and B and the normal reflection ($R^N$) coefficient at interface B. Interface A is set to have $R^N = 0$ to reflect its very transparent nature. A generalized Landauer-Büttiker formula is used to calculate the two-terminal conductance $G$ as a function of the incident carrier energy $E$. For energy $E$ above the Fermi surface, $G(E)$ is given by

$$G(E) = \frac{2e^2}{h} \sum_{i=1}^{M(E)} \int \frac{\partial f(\varepsilon - E)}{\partial E} \left(1 - R_i^N(\varepsilon) + R_i^A(\varepsilon)\right) d\varepsilon \quad (S3)$$

[1] where $M(E)$ is the number of transverse modes for the incident electron with energy $E$, and $f(\varepsilon)$ is the Fermi distribution function at temperature $T$. $\partial f(\varepsilon - E)/\partial E$ is replaced by a $\delta$-function at zero temperature.

We systematically varied parameters $\Delta_0$, $\Delta_{\text{BLG}}$, $\mu_{\text{BLG}}$, $T$, the width of region II $L_{\text{II}}$, and the location of the $p$-$n$/$p$-$p$ junction in the BLG to examine their impact on $G(E)$. Sections 2-5 in Appendix F describe considerations that are important in capturing the experimental features. They offer general insights on the behavior of van der Waals superconducting junctions.

**2. The necessity of a proximity-induced gap in BLG and its smooth decay at interface**

Our simulations show that the presence of a proximity-induced SC gap in region II is crucial to reproducing the dome plus peak/dip feature observed in experiments (Fig. 2). Figure 10 compares two scenarios where $\Delta_{BLG}$ is set to 0 (blue) and $0.2\Delta_0$ (red) respectively with the latter corresponding to the circles marked in Fig. 2(a). The red trace captures very well the curvature change observed in the hole data shown in Fig. 2(c), which is absent in the blue trace. The zero-bias conductance of the red trace doubles that of the normal state due to the absence of any normal backscattering. The calculated $G(E)$ for the electron regime develops a corresponding dip at $0.2\Delta_0$, as shown in Fig. 3(d).

We have also varied the decay length of $\Delta_{BLG}$ at the superconducting-normal BLG interface and found that a smooth decay is necessary to capture the zero-bias peak of the hole data exhibited in Fig. 3. The comparisons are shown in Fig. 11. An abrupt change of $\Delta_{BLG}$ leads to normal backscattering and conductance suppression at low energies due to the chemical potential mismatch present at the *p-p* junction [Fig. 11(a)].

**3. The location of the chemical potential variation in BLG and conductance oscillations**

In this section, we vary the location of the *p-n*/*p-p* junction with respect to $\Delta(x)$ to illustrate two points. Figure 10(a) illustrates the setup, where $\Delta(x)$ is fixed while $L'$ varies from 0 to $100a$. The computed $G(E)$ curves are shown in Fig. 12(b). A zero-bias conductance dip develops only with $L' = 80a$ and $100a$, i.e. when the chemical potential variation occurs at interface B or in the normal BLG region, which correspond to real device situations shown in Fig. 9. The physical picture is also intuitive. A *p-n* junction in the normal region of the BLG contributes to conductance suppression. In addition, pronounced conductance oscillations occur outside the central feature, similar to what is shown in Fig. 3(d). These oscillations are the Fabry-Perot interference effect occurring in region $L'$ between the Andreev reflection at interface A and the normal reflection at interface B'. It is much less pronounced in the hole regime due to a much smaller $R^N$ at interface B. Its period in $E$ decreases with increasing $L'$, as expected. In experiment, the vertical tunneling between BLG and $NbSe_2$ can occur at different locations so the interference effect is absent. In simulations, we average $G(E)$ computed with varying $L_{II}$. Conductance oscillations are effectively suppressed and the averaged results shown in Fig. 3(e) reach good agreement with data.

**4. The finite sample size effect**

As discussed in Sec. 1, the finite sample size in our simulation introduces a quantization energy splitting $\delta E \sim 0.01t$. Figures 13(a) and 13(b) show that as the SC gap in $NbSe_2$ $\Delta_0$ approaches this energy scale, the calculated conductance curves lose the "dome plus peak/dip" feature observed in experiment. The quantization effect is not important in experiment due to large device dimensions. Thus, we have chosen $\Delta_0 = 0.05t$ to eliminate the finite sample size effect in the simulations.

**5. The effect of finite temperature**

Finally, we examine the effect of temperature. As Figs. 14(a) and 14(b) show, the increase of $T$ leads to the rounding of the sharp features seen in the $T = 0$ simulations. In our experiment, $k_BT$ is approximately $0.1\Delta_0$. Indeed, the $k_BT = 0.1\Delta_0$ traces in Figs. 14 resemble measurements in Figs. 2 and 4 quite well.

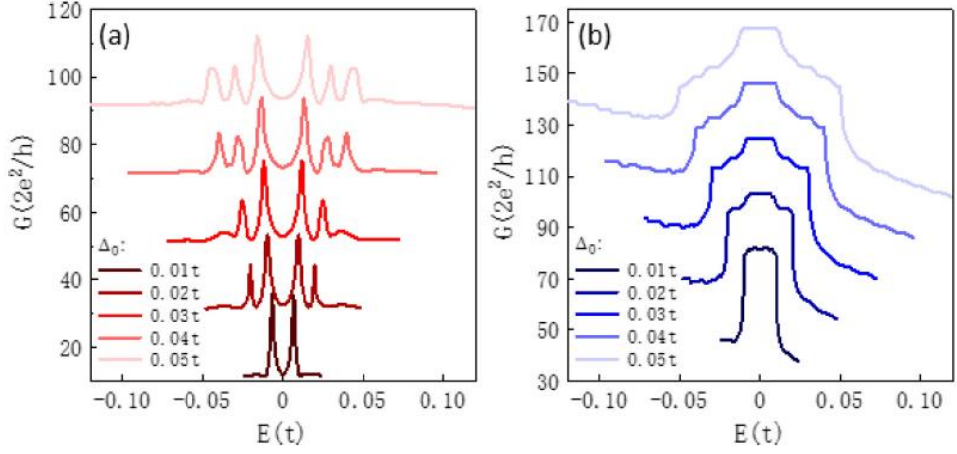

FIG. 13. (a) and (b) plot simulated *G* (*E*) in the electron and hole doping regime, respectively at selected $\Delta_0$ ranging from 0.01*t* to 0.05*t*. *T* = 0. See Fig. 3 for other parameters. Curves are vertically shifted by 20 units for clarity.

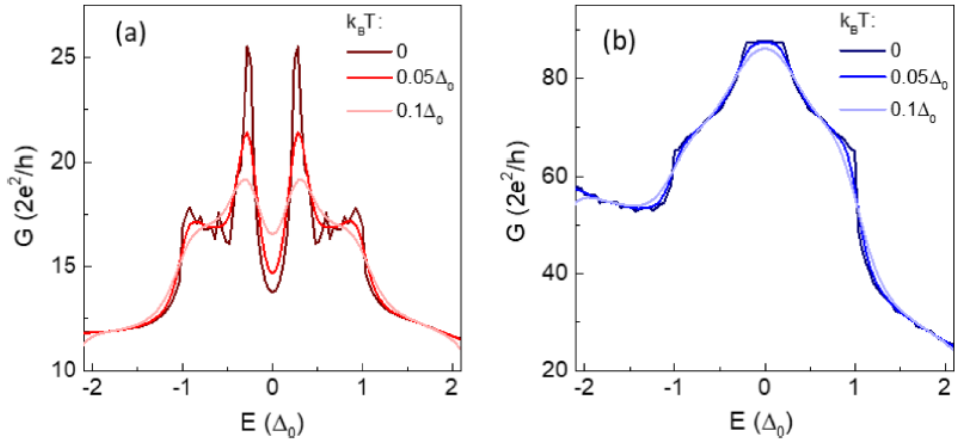

FIG. 14. Simulated *G* (*E*) in the electron (a) and hole (b) regime with varying temperatures as labeled. See Fig. 3 for other parameters.